\newcommand\Msun{\hbox{M$_\odot$}}

\newcommand\kms{\hbox{$\,$km$\,$s$^{-1}$}}
\newcommand\Apx{\hbox{$\,$\AA$\,$px$^{-1}$}}

\newcommand\one{\,{\sc i}}
\newcommand\two{\,{\sc ii}}
\newcommand\three{\,{\sc iii}}

\newcommand\tmult{\multicolumn{2}{c}}
\documentclass[12pt,preprint]{emulateapj}
\usepackage{times}
\usepackage{graphicx}
\shorttitle{Gemini Spectroscopy and HST Imaging of Region~B in M82}
\shortauthors{I. S. Konstantopoulos et al.}
\begin{document}
\title{Gemini Spectroscopy and HST Imaging of the Stellar Cluster Population in Region~B of M82\footnote{Based on observations obtained at the Gemini Observatory, which is operated by the Association of Universities for Research in Astronomy, Inc., under a cooperative agreement with the NSF on behalf of the Gemini partnership: the National Science Foundation (United States), the Science and Technology Facilities Council (United Kingdom), the National Research Council (Canada), CONICYT (Chile), the Australian Research Council (Australia), CNPq (Brazil) and CONICET (Argentina).}$\ ^,$\footnote{Based on observations made with the NASA/ESA Hubble Space 
Telescope, obtained [from the Data Archive] at the Space Telescope 
Science Institute, which is operated by the Association of Universities for 
Research in Astronomy, Inc., under NASA contract NAS5-26555. These observations are associated with program \#10853}}
\author{I. S. Konstantopoulos\altaffilmark{1}}\email{isk@star.ucl.ac.uk}
\author{N. Bastian\altaffilmark{1}}
\author{L. J. Smith\altaffilmark{2,1}}
\author{G. Trancho\altaffilmark{3,4}}
\author{M. S. Westmoquette\altaffilmark{1}}
\author{J. S. Gallagher III\altaffilmark{5}}

\altaffiltext{1}{Department of Physics and Astronomy, University College London, Gower Street, London, WC1E 6BT, UK}
\altaffiltext{2}{Space Telescope Science Institute and European Space Agency, 3700 San Martin Drive, Baltimore, MD 21218, USA}
\altaffiltext{3}{Universidad de La Laguna, Avenida Astr\'{o}fisico Francisco S\'{a}nchez s/n, 38206, La Laguna, Tenerife, Canary Islands, Spain}
\altaffiltext{4}{Gemini Observatory, 670 N. A'ahoku Place, Hilo, HI 96720, USA}
\altaffiltext{5}{Department of Astronomy, University of Wisconsin-Madison, 5534 Sterling, 475 North Charter Street, Madison, WI 53706, USA}
\clearpage
\begin{abstract}
We present new spectroscopic observations of the stellar cluster population of region~B in the prototype starburst galaxy M82 obtained with the Gillett Gemini-North 8.1-metre telescope. By coupling the spectroscopy with \emph{UBVI} photometry acquired with the Advanced Camera for Surveys (ACS) on the Hubble Space Telescope (HST), we derive ages, extinctions and radial velocities for seven young massive clusters (YMCs) in region B. We find the clusters to have ages between 70 and 200\,Myr and velocities in the range 230 to 350\kms, while extinctions  $A_V$ vary between $\sim$1--2.5 mag. We also find evidence of differential extinction across the faces of some clusters which hinders the photometric determination of ages and extinctions in these cases. The cluster radial velocities indicate that the clusters are located at different depths within the disk, and are on regular disk orbits. Our results overall contradict the findings of previous studies, where region~B was thought to be a bound region populated by intermediate-age clusters that formed in an independent, offset starburst episode that commenced 600\,Myr--1\,Gyr ago. Our findings instead suggest that region B is optically bright because of low extinction patches, and this allows us to view the cluster population of the inner M82 disk, which probably formed as a result of the last encounter with M81. This study forms part of a series of papers aimed at studying the cluster population of M82 using deep optical spectroscopy and multi-band photometry.
\end{abstract}
\keywords{galaxies: evolution --- galaxies: individual (M82) --- galaxies: photometry ---galaxies: spectroscopy --- galaxies: starburst --- galaxies: star clusters}

\clearpage
\section{Introduction}\label{intro}

The extensively studied galaxy M82 is a local example of a nuclear starburst galaxy. \citet[hereafter OM78]{om78} first catalogued the complex star-forming regions seen in ground-based images of the M82 disk, and introduced the nomenclature A--H. Region~B is the brightest region in the disk and is located 350--1050\,pc north-east of the nucleus. OM78 and \citet{marcum96} find that the integrated spectrum of this region is indicative of a fossil starburst region as it shows the characteristic `E$+$A' post-starburst spectrum, suggestive of a truncated burst of star formation occurring 100--1000\,Myr ago. Moreover, they find that the intrinsic brightness of region~B is such that the burst must have been of comparable intensity to the present starburst in the nucleus.

More recently, de Grijs, O'Connell, \& Gallagher (2001) studied the M82-B cluster population using photometry obtained with the Wide Field and Planetary Camera 2 (WFPC2) and the Near-Infrared Camera and Multi-Object Spectrometer (NICMOS) onboard the Hubble Space Telescope (HST). They identified 113 clusters and, by estimating ages and extinctions from $BVI$ photometry, they found that a concentrated episode of cluster formation occurred 400--1000\,Myr ago with a peak at 600\,Myr. Subsequent to this analysis, de Grijs, Bastian, \&~Lamers~(2003a) re-derived ages and extinctions by combining $BVI$ and $JH$ photometry; they find that the peak of cluster formation occurred at 1.10\,Gyr with an age spread of 500--1500\,Myr. de Grijs, Bastian, \&~Lamers~(2003b) used these new ages to derive the cluster luminosity function (CLF) for a fiducial age of 1.0\,Gyr and find that it has a Gaussian shape, similar to globular clusters (GCs), rather than the standard power-law CLF found for populations of young massive clusters (YMCs) \citep[e. g. ][]{larsen04,gieles06a}. This result is surprising but supports theoretical models advocating that an initial power-law distribution of cluster masses will be transformed into a Gaussian distribution because low mass clusters will be preferentially disrupted \citep[e.g.][]{fall01, vesperini03, gieles06b}. \citet{RdG03-2} therefore suggest that because M82-B is of intermediate age, it provides the `missing evolutionary link' between the power-law CLFs found for YMCs and the Gaussian CLFs of globular clusters. However, for an environment as turbulent as that of M82, one would expect a timescale of several Gyr for preferential disruption to produce this effect \citep[e. g. ][where such a timescale is calculated for NGC 1316 and NGC 3610 respectively]{goud1316,goud3610}.

To date, spectroscopy of the region~B cluster population has been extremely limited; the only published spectra are those of \citet{stis} for the two brightest members. The spectra were obtained with the Space Telescope Imaging Spectrograph (STIS) and, although they are of low quality, the derived ages of 350$\pm$100\,Myr are much lower than the photometrically-based ages of $\sim$0.7--6~Gyr \citep{stis, RdG03-1} for the same clusters. This discrepancy hints at the possibility that region~B may be younger and that the ages derived from the photometry are too high. It is important to determine accurate ages for the M82-B cluster population to verify its unusual CLF, and for understanding the cluster formation history of M82 and its relationship to encounters with its close neighbour M81.

 We therefore acquired new spectroscopic and photometric data for the M82-B cluster population. By using both techniques we are able to obtain information for a large number of clusters and also overcome the degeneracy between age and extinction that presents a hurdle in the analysis of photometric data \citep{gelys07a, gelys07b}. In this paper, we present optical spectroscopy for seven clusters obtained with the Gillett Gemini \mbox{8.1-m} telescope, and in a companion paper \citep{phot}, we present new HST imaging, including $U$-band photometry of the clusters.

\section{Observations and Data Reduction}\label{data}
\subsection{Spectroscopy}

\begin{figure*}
\begin{center}
\includegraphics[width=\textwidth]{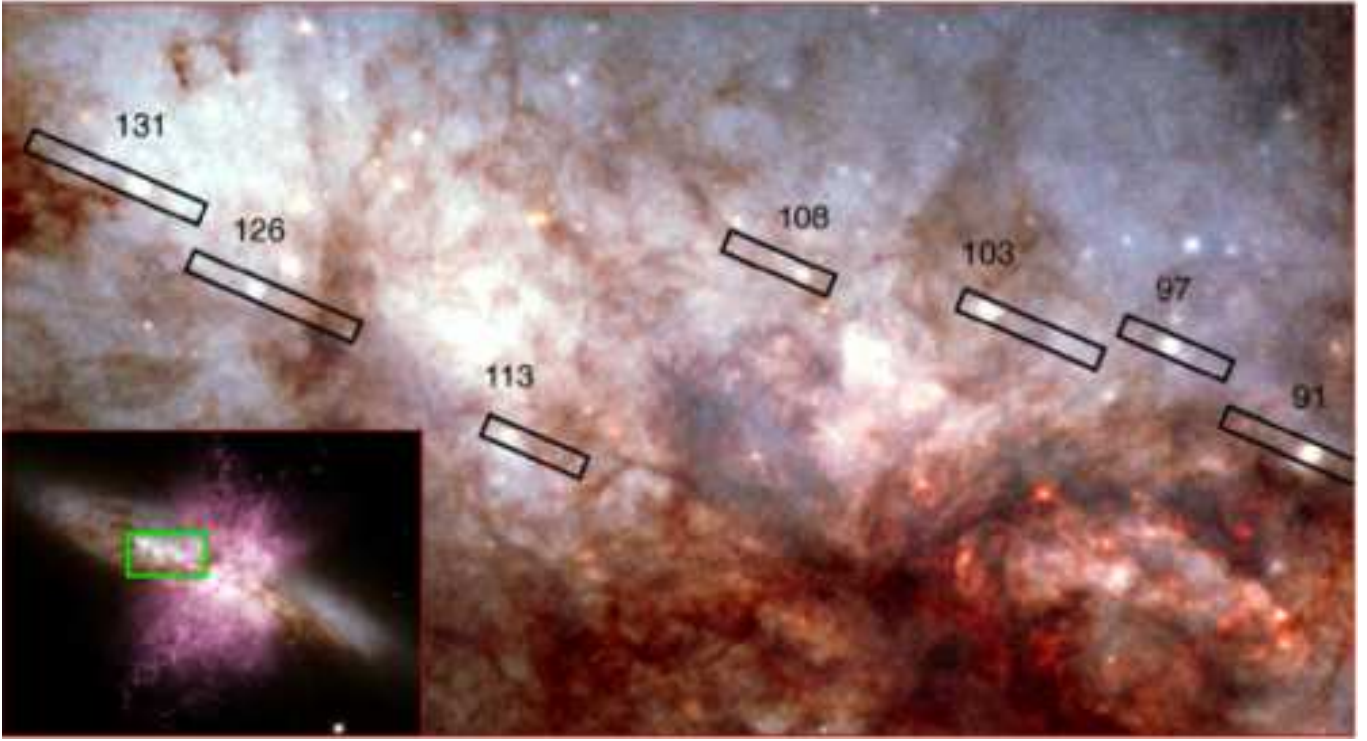}
\caption{Finding chart for all the clusters in region~B discussed in this paper. The inset shows the location of the region within the galaxy. East is towards the left and North is to the top of both images. The rectangles represent the slits used to acquire the data, to scale, with identifiers placed above the clusters. We choose to study these seven clusters because they are bright and isolated. Note the intricate structure of the gas and dust in this region. Both images are composites of four filters, $B$, $V$, $I$ and $H\alpha$ \citep[taken from the Hubble Heritage HST ACS mosaic; see][]{mosaic}.}
\label{plate}
\end{center}
\end{figure*}

\begin{figure*}
\begin{center}
\includegraphics[width=\textwidth]{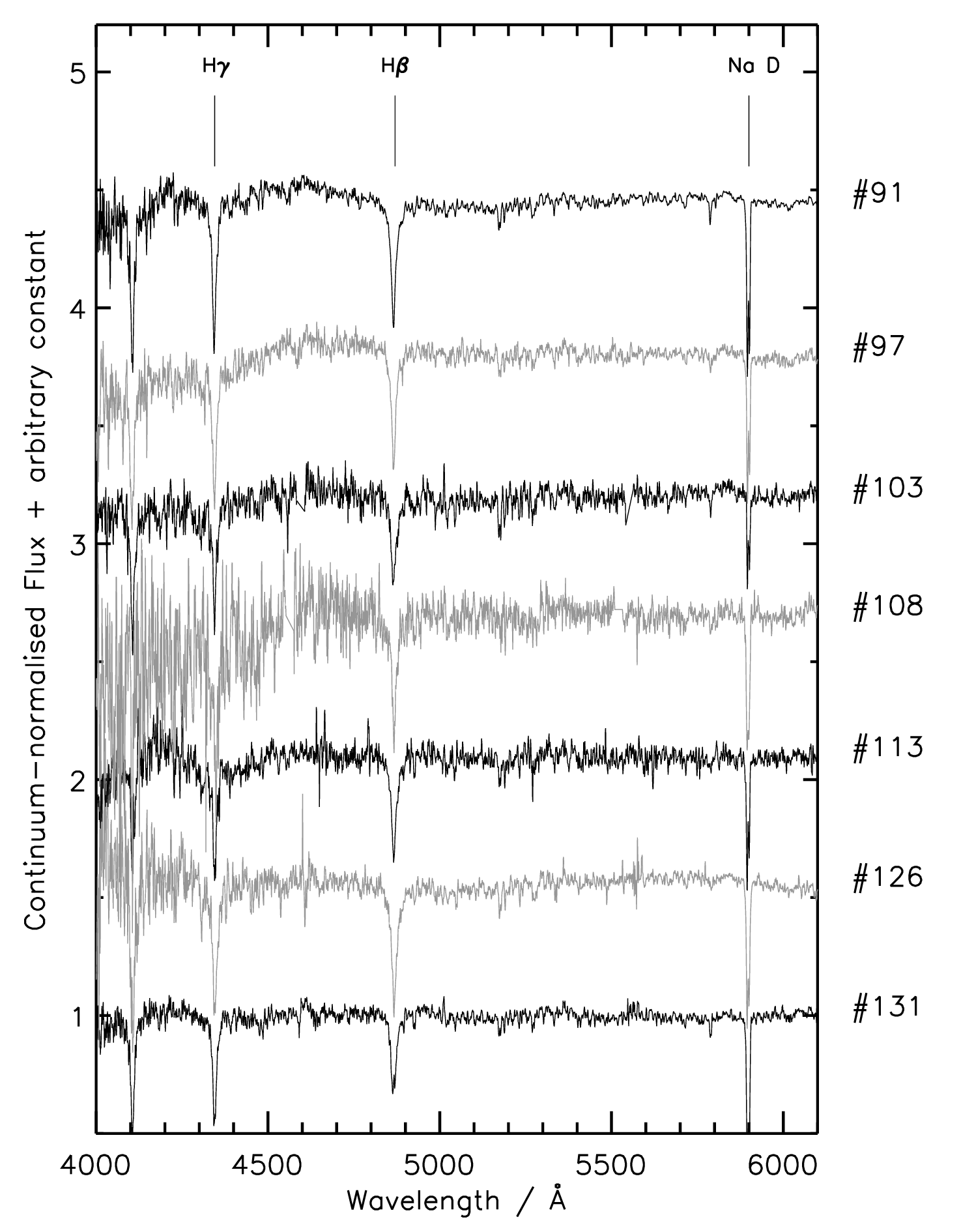}
\caption{Continuum-normalised Gemini spectra of the seven clusters presented in this paper. The wavelength range is 4000-6100\,\AA\ for all spectra. We use an arbitrary flux-scale for means of presentation, however, all analysis was performed on absolutely flux-calibrated data in the presented wavelength range. The locations of the lines we use as our main diagnostic tools are indicated and labelled at the top of the plot.}
\label{spectra}
\end{center}
\end{figure*}

The spectroscopic data presented in this paper were obtained with the Gillett Gemini North 8.1-metre telescope using the Gemini Multi-Object Spectrograph (GMOS-N), as part of program GN-2006A-Q-38 (PI Smith). The target selection was made based on photometric criteria with the potential YMCs identified on GMOS pre-imaging, taken as part of the observing program. A multi-slit mask was designed, using a slit width of 0$\farcs$75, with a variable slit length that we tried to keep to a minimum of 6$\farcs$5, in order to include an area of sky that would ensure a good background subtraction. In some cases however, a smaller slit length was used so as to include a greater number of sources on the mask, resulting in a total of 39 objects being targeted across the M82 galactic disk. In this paper we focus on seven of these 39 clusters that are situated within region B. Fig.~\ref{plate} indicates the positions of the clusters and the slits, and the fully reduced rectified spectra are presented in Fig.~\ref{spectra}. A list of cluster identifiers and coordinates can be found in Table~\ref{tab-radec}.

The data were acquired on the night of 2006 April 5, under good seeing\footnote{This is not the seeing, but the `image quality' parameter adopted by Gemini, which incorporates both wind and telescope performance effects.} conditions (0$\farcs$8 at 5000\,\AA).  We used the B600 grating and the data were read-out in 2$\times$2 binning mode, resulting in a dispersion of 0.9\Apx. As this is multi-object spectroscopy, the central wavelength varies slightly with every slit, so the resulting spectral range of the data is approximately~3700-6500\,\AA. While we used all three GMOS-N CCDs to achieve this broad range, the sensitivity of the silver-mirrored chips is lower at the blue end, resulting in a poor S/N and loss of accuracy in the wavelength calibration below 4000\,\AA, thus setting a lower limit to our effective wavelength range.

The data were acquired in eight 1800 second exposures, totalling four hours of integration time. More specifically, there were two sets of four exposures that differed in terms of central wavelength. This is standard GMOS procedure which allows the observer to compensate for the inter-chip gaps through co-adding/merging the spectra; the two $\lambda_{\rm central}$ were 5080 and 5120\,\AA. CuAr (Copper-Argon) arc exposures and Quartz-Halogen flat fields were taken in between target exposures. Bias frames were taken as part of the Gemini base-line calibrations (GCAL).

The data were subsequently reduced using mainly standard IRAF\footnote{IRAF is distributed by the National Optical Astronomical Observatories, which are operated by the Association of Universities for Research in Astronomy, Inc. under contract with the National Science Foundation} reduction tools, as well as tasks from the purpose-developed Gemini-IRAF package (v1.9.0). The data were bias-subtracted, flat-fielded and then combined into a single image, while correcting for discrepancies in quantum efficiency between the three detectors using the QE Gemini IRAF tool. The wavelength calibration was performed using the obtained CuAr arc frames, resulting in residuals typically less than 0.15\,\AA. 

One complication that arises when observing in multi-slit mode without an active atmospheric diffraction correction device is differential refraction; this causes wavelength dependent slit-losses, as it is not possible to keep the slits aligned with the parallactic angle for the duration of an exposure. To alleviate this effect, the combined data were corrected for parallactic angle using a purpose-built IDL (Interactive Data Language) routine written by B.~W.~Miller~(2006, priv. comm.), based on \citet{filippenko}, where the effects of atmospheric differential refraction are discussed.

At this point we extracted all 39 spectra from each multi-dimensional MOS frame and merged all available exposures of each target. In most cases we decided not to use all available spectra, as some exposures were of poor quality (caused by seeing degradation) and thus lowered the quality of the merged spectrum.

Finally, the spectra were flux-calibrated using a response function derived from the standard star Wolf 1346. It should be noted here that the observations used were taken on a different night (2006 May 25), due to problems that arose with the exposure of the original flux standard, Feige 34. This unfortunately led to an increased uncertainty in the absolute flux scale.

Having completed the extraction procedure, we measured the width of the sky-line at $\lambda$5577, in order to measure the spectral resolution of our data, which we found to be 3.5\,\AA~(FWHM). We also tested the wavelength calibration by measuring the exact position of this line and found it to be well within the predicted errors. The signal-to-noise ratios (S/N) of the spectra were determined using two continuum windows between 4400-4800\,\AA\ and 5000-5800\,\AA. We found values ranging from 16--44; the individual values for each cluster are presented in Table~\ref{tab-spec}.

\begin{table}
\caption{Positional information for the cluster sample.\label{tab-radec}}
\begin{center}
\begin{tabular}{lllcc}
\tableline
\tableline
 \#\tablenotemark{a} & Alt \#\tablenotemark{b} & Previous \#\tablenotemark{c} & RA\tablenotemark{d} & DEC\tablenotemark{e}\\
\tableline
91	&	34	& B2-12, H\tablenotemark{f}, B1-2\tablenotemark{g}  &   9\ 55\ 54.60	&	69\ 41\ 01.7\\
97	&	25	&	B2-26  					&   9\ 55\ 55.65	&	69\ 41\ 05.9\\
103	&	20	&	B2-41  					&   9\ 55\ 57.04	&	69\ 41\ 07.4\\
108	&   \ldots	&	B2-49					&   9\ 55\ 58.41	&	69\ 41\ 08.7\\
113	&	14	&	B1-6						&   9\ 56\ 00.63	&	69\ 41\ 02.5\\
126	&	6	&	B1-20					&   9\ 56\ 02.52	&	69\ 41\ 08.1\\
131~~ &	1	&	B1-28, B2-1\tablenotemark{h}	&   9\ 56\ 03.40	&	69\ 41\ 12.3\\
\tableline
\end{tabular}
\end{center}
\tablenotetext{a}{Cluster identifier as in our Gemini program}
\tablenotetext{b}{Identifier as in \citet[][]{phot}.}
\tablenotetext{c}{Identification by \citet[][unless otherwise stated]{RdG01}, with `B1' and `B2' referring to the part of the region to which the cluster belongs.}
\tablenotetext{d,e}{J2000 coordinates as measured on the F555W ACS mosaic image; the units are (h m s) for $\alpha$ and ($^{\circ}$ $\prime$ $\prime\prime$) for $\delta$.}
\tablenotetext{f}{Notation follows OM78.}
\tablenotetext{g,h}{After STIS study of \citet{stis}.}
\end{table}

\subsection{Photometry}
\label{data_im}

We used imaging from the recent Hubble Heritage ACS Mosaic of M82 \citep[for full details of the observations see][]{mosaic}. This is a composite of six individual pointings, resulting in a combined field of view of 12288$\times$12288 pixels, or approximately 9$\times$9 arcminutes squared, given the 0.05 arcsec per pixel scale of the ACS-WFC. The images cover the entire galaxy in four bands, of which we use the F435W, F555W, and F814W that correspond roughly to the Johnson \emph{B}, \emph{V} and \emph{I}. We performed aperture photometry of the sources in region B using an aperture of 10~pixels (0$\farcs$5) and a background annulus with inner radius and width of 11 and 1~pixels respectively.

We also used \emph{U}-band (F330W) ACS-HRC imaging of region~B, obtained as part of HST program \#10853 (PI Smith), which extends our photometric coverage to the near-UV. The combination of optical and UV data enables us to set a strong constraint on age and extinction determinations of young stellar clusters \citep[e.g.][]{anders04}.  The HRC data were processed with the standard automatic HST pipeline and drizzled (using the MultiDrizzle package -- Koekemoer et al.~2002) to a scale of 0$\farcs$025 per pixel. For a more detailed discussion of the data we refer the reader to \citet{phot}. We also performed aperture photometry on the HRC images with apertures of the same physical size (0$\farcs$5 as projected onto M82), that is, an aperture radius of 20~pixels and a background annulus of inner radius and width of 22 and 2 pixels respectively.

For both the WFC and HRC data, we transformed the measured flux to the {\sc vegamag} system using the zero-points given in \citet{sirianni05}.

Due to the extended nature of the clusters in this study, we derived aperture corrections for each of the images based on bright, isolated clusters in our field of view (namely clusters \#~91 and 97).  The corrections were found by comparing the cluster magnitudes as measured by applying apertures of two different radii (10 and 40 pixels for the WFC images and 20 and 80 pixels for the HRC images).  The applied corrections are $-$0.64, $-$0.35, $-$0.33 and $-$0.43 magnitudes in the $U$, $B$, $V$~and~$I$ bands respectively.  We note, however, that since the size and profile of young clusters vary \citep[e.g.][]{larsen04, bastian05} these corrections may not apply precisely to all clusters studied, although we expect that the different cluster sizes will not affect the derived colours.

Note that throughout the paper we use the Johnson $UBVI$ notation for sake of simplicity, but we do not convert magnitudes to this system.

\section{Analysis and Results}\label{analysis}

The combination of {\it UBVI} photometry and high S/N optical spectra allows us to determine the age and extinction of each of the seven clusters in our sample.  Throughout this section we assume that the clusters have solar metallicity, in agreement with ISM studies of M82 \citep{mcleod93}. Additionally, we adopt a standard Galactic extinction law \citep{savage79}.

We note here that the assumptions on the metallicity and extinction laws will not have large impacts on our conclusions. In particular, if we were to use the starburst extinction-law of \citet{calzetti97} the photometric ages and extinctions would remain virtually unchanged.

\subsection{Ages and extinctions from photometry}
\label{sec:ages_phot}

\begin{table}
\begin{center}
\footnotesize
\caption{Summary of photometrically derived results.}
\begin{tabular}{@{}lcccccccr@{, }l@{}}
\tableline
\tableline
 & F330W & F435W & F555W & F814W & ~$M_V$ & $A_V$ & Age & \tmult{$\sigma_{\rm age}$\tablenotemark{a}} \\
 \# & (mag) & (mag) & (mag) & (mag) & (mag) & (F555W) & (Myr) & \tmult{(Myr)}\\
\tableline
91	& 20.12 & 18.91 & 17.75	& 16.07	& $-12.95$	& 2.4\tablenotemark{b} & \ldots\tablenotemark{c} & \tmult{\ldots}\\
97	& 18.93 & 18.81 & 18.28	& 17.25	& $-10.56$	& 1.1 	& 180	&	[140&220]\\
103	& 19.21 & 19.08 & 18.35	& 16.94	& $-11.36$	& 1.9	& \phantom{1}90		&	[60&150]\\
108	&  \ldots\tablenotemark{d}	& 20.04 & 19.21 & 17.84 & $-12.24$	& 2.5\tablenotemark{e} & \ldots\tablenotemark{f} &\tmult{\ldots}\\
113	& 20.41 & 20.10 & 19.35 	& 18.01	& $-10.30$	& 1.9 	& 170	&	[72&360]\\
126	& 19.77 & 19.34 & 18.64 	& 17.47 	& $-10.70$	& 1.5 	& 260	&	[190&1200]\\
131~~& 18.37 & 18.24 & 17.74	& 16.61	& $-11.22$	& 1.2 	& 170	&	[110&190]\\
\tableline
\end{tabular}
\tablenotetext{a}{The minimum and maximum acceptable values for the cluster age according to the 3DEF method.}
\tablenotetext{b}{Value estimated from the extinction map in Fig.~\ref{h-ext}. This cluster is discussed in \S~\ref{sec:H}.}
\tablenotetext{c,f}{Not possible to obtain a photometric age because of differential extinction (see \S~\ref{sec:H}).}
\tablenotetext{d}{Cluster is not visible in the $U$-band.}
\tablenotetext{e}{Estimated from BVI photometry assuming spectroscopically derived age.}
\label{tab-phot}
\end{center}
\end{table}

As a first method to determine the ages and extinctions of the clusters we compare their colours to simple stellar population (SSP) models. We choose the GALEV SSP models \citep{anders03}, with solar metallicity and a Salpeter IMF.  These models use the Padova stellar isochrones, and the evolution of magnitude is given as a function of time directly in the {\sc vegamag} system for HST filters. In that way there is no need to convert the measured cluster magnitudes to the more standard Cousins-Johnson system, which would introduce additional errors.

\begin{figure}
\begin{center}
\includegraphics[width=0.49\textwidth]{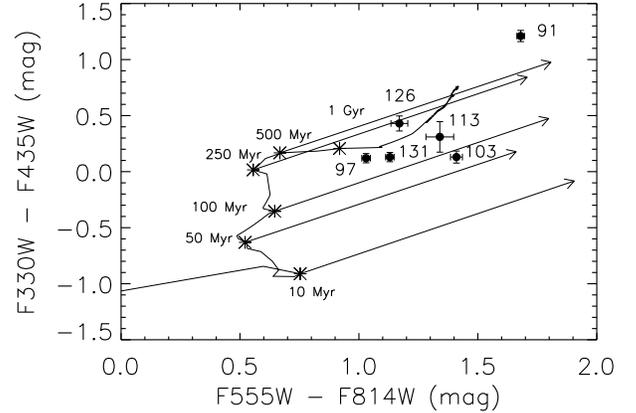}
\caption{Colour-colour plot for six of the seven clusters in \mbox{M82-B}. The solid line traces the GALEV SSP model colours, with asterisks marking model track ages of 10, 50, 100, 250, 500 and 1000\,Myr. The data-points represent the cluster colours and associated errors, as derived from multi-band photometry. The arrows represent extinction vectors, along which the `footprint' of the cluster may be traced, in order to estimate its age and extinction; the length of the vectors is 3 magnitudes. The large $V-I$ error on cluster \#113 is due to contamination from a neighbouring red source. The errant cluster \#91 has an underestimated \emph{U}-band flux (see Sect. 3.4) and cluster \#108  is not included because it is not detected in the \emph{U}-band.}
\label{fig:colplot}
\end{center}
\end{figure}

In Fig.~\ref{fig:colplot} we show the $U-B$ vs. $V-I$ colours of the clusters in region~B for which we have UBVI photometry and spectra.   Additionally, we demonstrate the evolution of the GALEV SSP models by labelling six ages, and we show extinction vectors of magnitude $A_V=$3.0 starting from these ages to guide the eye.  From this plot we can deduce that all but one of the clusters (\#91) have ages between 50\,Myr and 1\,Gyr.  Cluster \#91 (cluster H, following notation by OM78) will be discussed in \S~\ref{sec:H}.

However, as the SSP colours do not evolve monotonically in colour-space, the $U-B$ vs. $V-I$ diagram provides simply one of the possible colour projections of the photometry.  In order to fully exploit the available data, we have used the 3DEF method presented in \citet{bik03} and further tested in \citet{bastian05}.  Briefly, this method compares the photometry of each cluster to that of the SSP models, to which it applies an extinction ($A_V$) between 0 and 5 mag, in steps of $\Delta A_V=0.02\,$mag and calculates a $\chi_{\nu}^2$ (reduced $\chi^2$).  The model, i.e. the combination of age and extinction with the lowest $\chi_{\nu}^2$ is selected as the best fit and the errors are assigned by the extrema of the set of models which satisfies the condition $\chi_{\nu}^2 < \chi_{\nu,\rm best}^2 + 1$.

The photometrically derived best fit ages range between 90 and 260\,Myr and we find $A_V$ extinction values in the range 1.1--2.5. The values for each cluster are shown in Table~\ref{tab-phot}.

\subsection{Spectroscopy}\label{sec:ages_spec}

\begin{table}
\caption{Spectroscopically derived age ($\tau$) and radial velocity ($u$) measurements (heliocentric).\label{tab-spec}}
\begin{center}
\begin{tabular}{lcccccr@{, }l}
\tableline\tableline
 & S/N\tablenotemark{a} & $v_R$ & $v_R$ \scriptsize{(Na\one)}\tablenotemark{b} & $\tau_{\rm MSRM}$\tablenotemark{c} & $\tau_{\rm CCM}\,$\tablenotemark{d} & \tmult{$\sigma_\tau$\tablenotemark{e}}\\
\# & (\ldots) & \tmult{(\kms)} & \multicolumn{4}{c}{(Myr)}\\
\tableline
91	&	44	& $230\pm20$ 	& $270\pm10$	&	180	&	190	&	[110&250]	\\
97	&	34	& $290\pm20$ 	& $285\pm10$	&	180	&	200	&	[130&320] \\
103	&	23	& $270\pm20$ 	& $270\pm10$	&	710	&	100	&	[30&180]	\\
108	&	16	& $340\pm40$ 	& $200\pm30$ &	140	&	140	&	[60&250]\\
113	&	24	& $280\pm20$	& $225\pm10$	&	100	&	100	&	[20&180]\\
126	&	24	& $350\pm30$	& $220\pm20$ &	200	&	200	&	[130&280]\\
131~~&	40	& $260\pm10$ 	& $245\pm10$	&\phantom{1}80 & \phantom{1}70 & [40&160]\\
\tableline
\end{tabular}
\end{center}
\tablenotetext{a}{The average of calculations for two different continuum regions, $\lambda$4400-4800 and $\lambda$5000-5800.}
\tablenotetext{b}{The errors quoted here are adopted as a 0.25\,px measurement error, which roughly corresponds to 10\kms. In all but two cases the agreement between the two individual doublet line measurements gives rise to an error that is lower than the adopted systematic error.}
\tablenotetext{c,d}{The model-spectrum residual MSRM and cumulative CCM best fit ages (as explained in \S\,\ref{sec:ages_spec}).}
\tablenotetext{e}{The minimum and maximum acceptable ages according to the CCM.}
\end{table}

While photometry presents a useful tool for constraining ages and extinctions, its limitations (such as the need for detection in four filters spanning 5000\,\AA) show even in a sample of seven clusters. Spectroscopy has long been recognised as providing the most accurate means of determining cluster ages using SSP models. We have measured ages for our cluster sample using the solar metallicity models of \citet[][hereafter GD05]{gd05} in the age range \mbox{4\,Myr--1\,Gyr}, which are designed specifically for young clusters.  The parameters of the models match those assumed for the photometric fitting (solar metallicity, Salpeter IMF, and Padova stellar isochrones), and their wavelength range of 3000-7000\,\AA\ is well matched with that of our observed spectra. All model spectra have been degraded to match the observed spectral resolution of 3.5\,\AA.

We prefer using these theoretical models over the more established \citet{bc03} templates for a number of reasons. First the good age resolution means cluster ages can be determined with higher precision. Second, the high sampling resolution of 0.3\,\AA\  allows us to downgrade the resolution of the models, rather than that of the data, in order to match the two.

\begin{figure*}
\begin{center}
\includegraphics[width=0.9\textwidth]{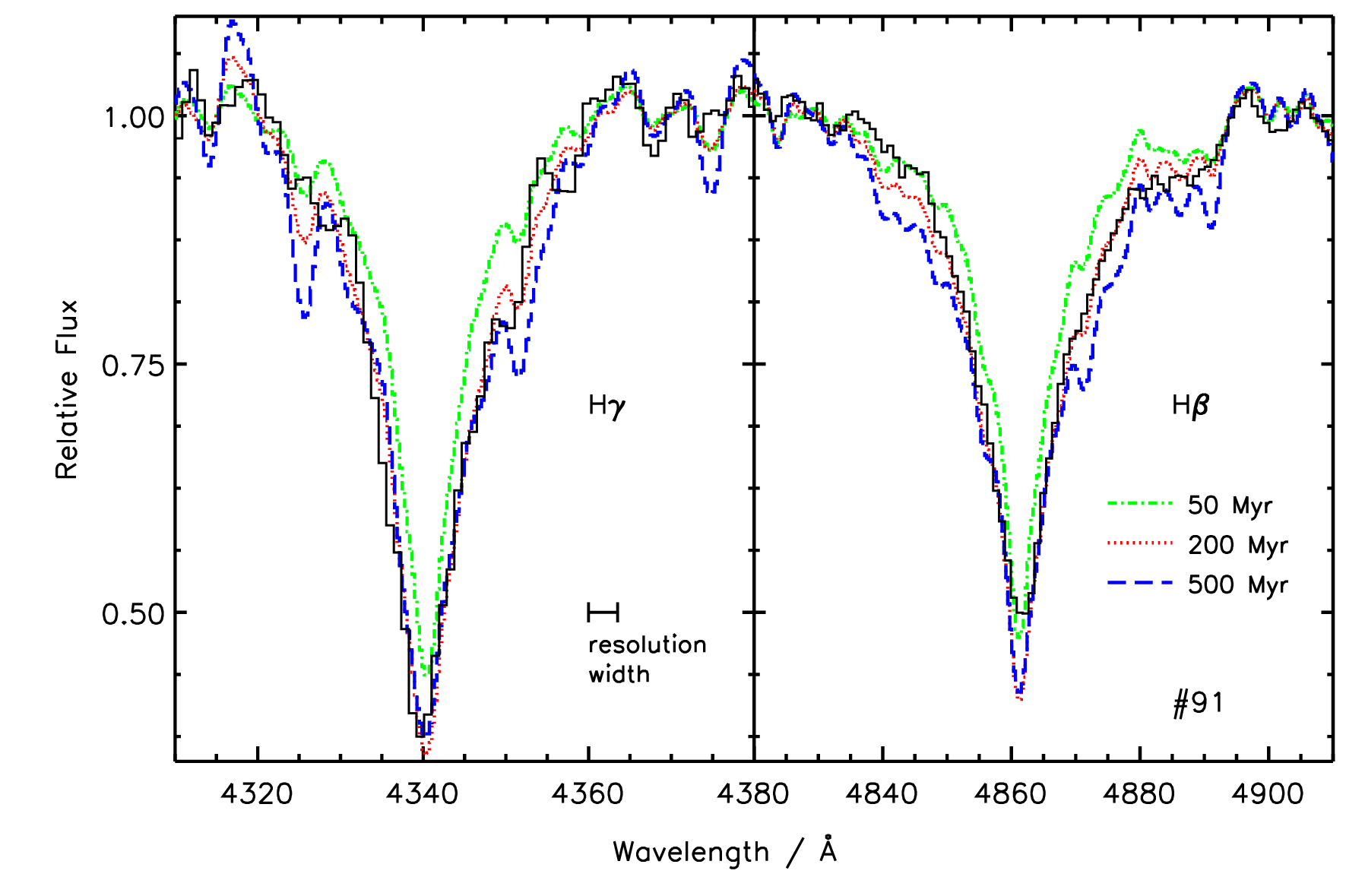}
\includegraphics[width=0.9\textwidth]{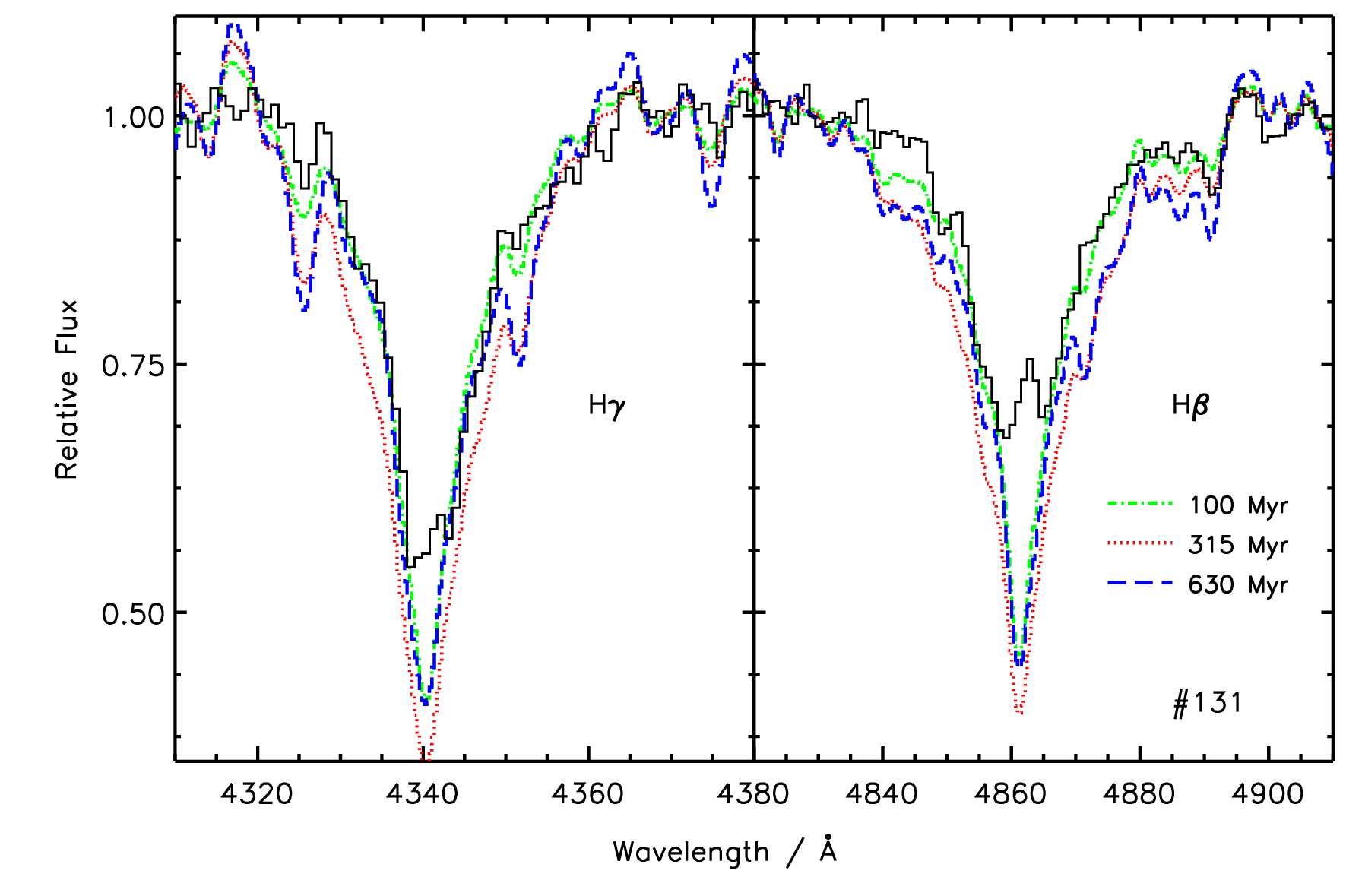}
\caption{H$\gamma$ and H$\beta$ line fitting for two of the clusters in region~B. The over-plotted models have been chosen to demonstrate the variation of Balmer line profiles with age and the effect on the fit to an observed spectrum. {\bf The top panel} shows cluster \#91, which stands out in Fig.~\ref{fig:colplot}, as it fails to meet the model track. We discuss the possible reasons for this in \S~\ref{sec:H}. A first inspection of this cluster might offer an age estimate as young as 50\,Myr, however, care must be taken as there is blended, embedded, off-set emission from the surrounding gas. Comparing the resolution width to that of the instrumental sampling, it is obvious that the Balmer lines are well resolved. {\bf The bottom panel} shows the line fits for cluster \#131, which we believe to have an age of $\sim$80\,Myr. Note the slightly red-shifted, strong superposed emission.}
\label{balmer}
\end{center}
\end{figure*}

We determine the ages of the clusters using two separate statistical methods, a model-spectrum residual method (hereafter MSRM) and a cumulative $\chi^2$ method (CCM). We choose for these clusters to investigate the fit of the models on the lower Balmer series spectral lines.  We compare the models and observations directly by first rectifying the spectra based on fits to the continuum in two windows outside of each line.  Fig.~\ref{balmer} demonstrates the Balmer line fits as applied to two of the clusters. By fitting both H$\beta$ and H$\gamma$, we can avoid having to make any presumptions about the shape of the line in cases where there is noise or nebular emission.  We use many aspects of the feature in order to determine the best fit; we check the agreement in line width, how well the overall line profile is fit and how closely the depth of the line is fit or predicted.  In most of the clusters we find emission superposed on the absorption line.  This emission presumably comes from diffuse ionised gas in the region and is not centred at the same wavelength as the absorption, which means that it is not directly associated with the clusters.

The main obstacle that arises when using automated statistical methods is the age degeneracy of Balmer line strength, i.e. multiple models of very different ages have similar line strengths. This means that the ages cannot be determined by simply finding the model that best predicts the depth of the line. In addition to the age degeneracy, the observed spectra may also feature superposed emission, so the fit would not in such cases reflect the actual absorption line profile.

The first method employed (MSRM) fits a model and calculates the minimised absolute value of the model-spectrum residual. The best fit is then chosen as the model with the lowest mean residual. The second method (CCM) uses all available data-points to calculate the $\chi^2$ cumulatively for each model fit. This offers a more realistic interpretation of a good fit to the entire line profile. In that way, the best-fitting template will be the one that traces the Balmer line profile, including the wings, thus overcoming the observed line strength degeneracy.

\begin{figure}
\begin{center}
\includegraphics[width=0.48\textwidth]{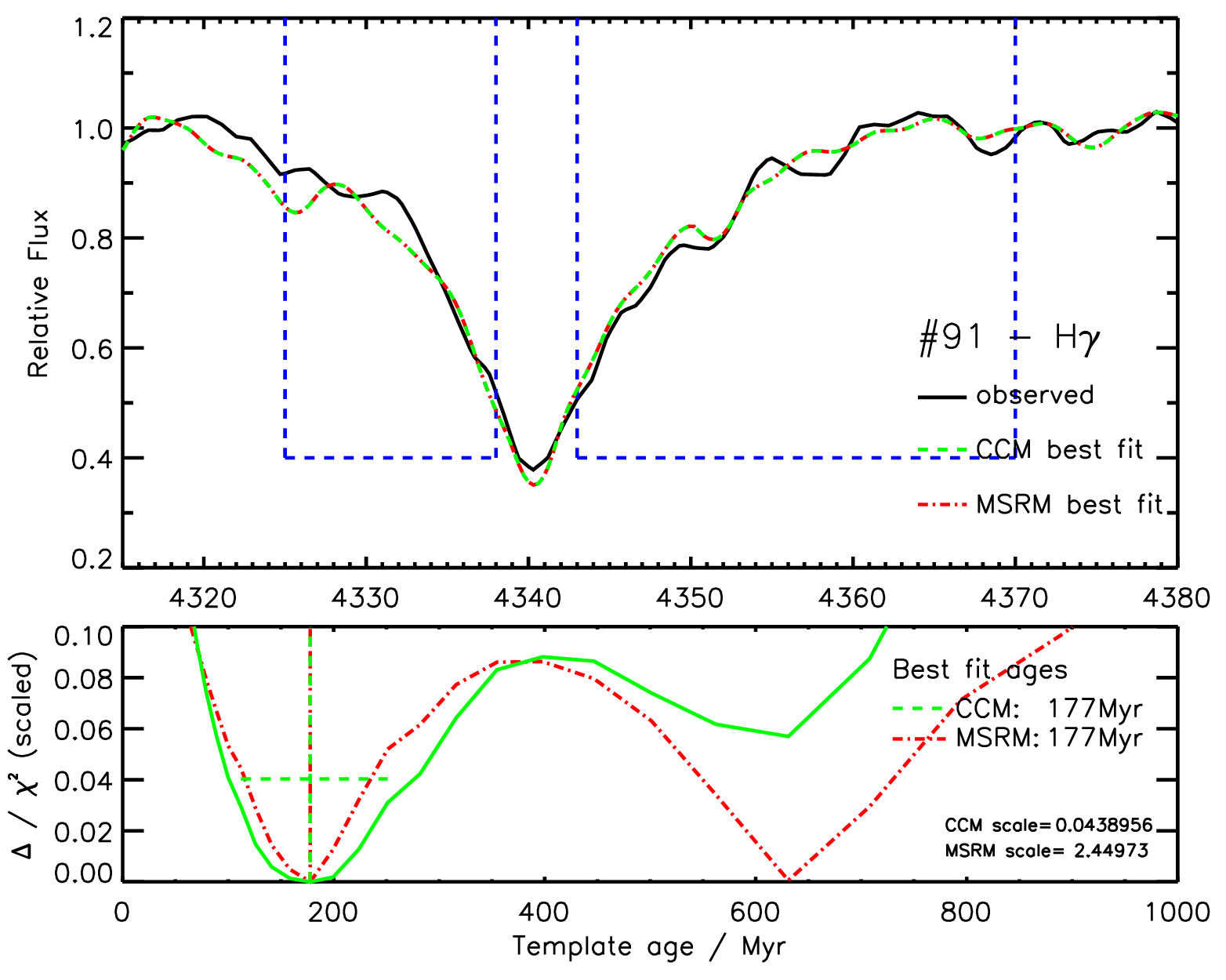}
\caption{Statistical fits for cluster \#97. The top panel shows the region of the observed spectrum in black (wavelength in~\AA) with the best fit models over-plotted; the dashed green and dashed-dotted red lines correspond to the CCM (cumulative $\chi^2$ method) and MSRM (model-spectrum residual method) best fits respectively. The blue boxes represent the regions of the spectrum considered for the fit. The bottom panel shows the statistical measures as a function of age; the solid green line traces the $\chi^2$ and the dashed-dotted red line the absolute value of the model-spectrum residual, with the vertical lines denoting the best fit model age for each method. In the case of cluster \#97, this occurs at 200 and 180\,Myr for the CCM and MSRM respectively (this applies to the H$\gamma$ fit). The horizontal dashed green line denotes the limit of values within $2\times\chi^2_{min}$ that we use as an error estimate. The y-axis traces the model-spectrum residual, and the $\chi^2$ curve has been scaled to 1/20 of its original values in order to fit in the same plot.}
\label{chi2}
\end{center}
\end{figure}

As mentioned above, most of the clusters contain emission features within the Balmer absorption lines.  Therefore, in fitting the data we have carefully avoided regions that are contaminated by emission.  This is shown in Fig.~\ref{chi2}, where the top panel shows the H$\gamma$ line for cluster \#97 (black solid line) and the two dashed boxes show where the fit was carried out. We would like to note here that including the profile centre in the fit causes degradation of the fit quality, however it leaves the cluster age distribution largely unaffected, with ages shifting by an average of $\sim$20~Myr towards younger values. The bottom panel of Fig.~\ref{chi2} shows the results of the two fitting methods applied as a function of age. The green (solid) line shows the  CCM and the red (dashed-dotted) line shows the  MSRM and both fits are carried out within the designated boxes. The  MSRM is more sensitive to line strength, and has two minima for very different ages, whereas the  CCM overcomes this obstacle in most cases and has one clear minimum.

The error on the best fitting age is calculated in a similar way as for the 3DEF method, that is, based on the extrema of the set of models which satisfies the condition~\mbox{$\chi^2 < 2 \times \chi_{\rm min}^2$}. We note here that a second $\chi^2$ minimum does tend to appear in the lower S/N spectra (clusters \#103, 108, 113 and 126). In these cases the very bottom of the second minimum `trough' is low enough to satisfy the mentioned error condition. Nonetheless we are able to confidently reject the higher age values by inspecting the resulting spectrum-model fit, as it shows a clear disagreement in the overall line profile. The derived ages and errors are given in Table~\ref{tab-spec}.

For the brightest cluster \#91, it is possible to use a third technique to further reinforce our results, the ratio of He\one, $\lambda$4471 to Mg\two, $\lambda$4481; by comparing the ratio between spectrum and model we can get one more estimate of the age of this cluster. This ratio is used widely as a spectral type determination method in stellar astrophysics \citep{walborn72,walborn90} and it provides an entirely extinction-independent age determination method (because of the proximity of the two lines). In order to overcome sampling-related errors that may arise when simply detecting the line minimum, we extrapolate the minimum by fitting Gaussian curves. This method has been successfully applied to the significantly reddened cluster M82-F \citep{bastian07}. Using this technique on cluster \#91, we obtain an age estimate of 180~Myr, which is in accord with all other age measurements.

The spectroscopic age measurements are summarised in Table~\ref{tab-spec}. In all, we find ages in the range 80--200\,Myr. We have adopted an age for each cluster as the average of all available spectroscopic age determinations that provide a good fit. The ages are listed in Table~\ref{tab-summ}. These results disagree with work previously published on region~B \citep{RdG01, RdG03-2}, where it was suggested that the ages of most clusters in the region are in excess of 400\,Myr. In the \citet{RdG03-2} sample of 113 clusters, there are only four with derived ages younger than 400 Myr. This age difference and its implications are discussed in more detail in  \S~\ref{res-age}.

\subsection{Radial Velocities}
\label{sec:rvs}
The clusters observed reside in a region rich in diffuse ionised gas. As a result, there is a small amount of Balmer line emission from their environment and therefore the Balmer line profiles of the clusters are a blend of the deep stellar absorption lines with narrow nebular emission lines, which in some cases are off-set. This makes the precise determination of the profile centre by inspection of Balmer lines practically impossible, therefore we resort to automated routines which also focus on other features to calculate the radial velocities of the clusters. To determine the velocity of the neutral gas in our line of sight, we use the Na\one\ D interstellar lines.

In order to derive the best possible estimates for the radial velocities we used a number of measurements (three to six measurements depending on the S/N), each one for a different spectral region and combined all results to get a value for the velocity (the arithmetic mean) and its error (the standard deviation).

The first method employed was the FXCOR cross-correlating routine in IRAF, which we applied to various regions of the spectrum, each time including a different set of absorption lines. We also used the Penalized Pixel Fitting Method (pPXF), taking care not to include the emission component of the Balmer line profiles in the fit. We will not provide a detailed description of this method here; instead, we refer the reader to \citet{ppxf}. In brief, the method simply cross-correlates selected regions of a spectrum with a template in order to derive the kinematics of the observed object. A significant benefit of using this routine is that it can fit a polynomial function to the overall shape of the template, thus matching the effect of extinction and differential refraction on the observed spectra. This enables us to avoid making corrections or normalising the spectra before fitting the templates. It also means that only one measurement is required, thus it does not introduce extra degrees of freedom in our calculation.

We find the two methods to provide consistent results when given the same templates to fit, resulting in accurate and precise radial velocity measurements. Correcting the cluster velocities to the heliocentric frame of reference gives values ranging between 230 and 350\kms\ (see Table~\ref{tab-spec}). 
This range corresponds to 30--150\kms\ with respect to the centre of the galaxy where we adopt a systemic velocity of 200\kms\ for M82 \citep{mckeith93}. The results are discussed in \S~\ref{sec:res-rvs}.

\subsection{Differential extinction on clusters \#91 and \#108}
\label{sec:H}
\begin{figure}
\begin{center}
\includegraphics[width=0.24\textwidth]{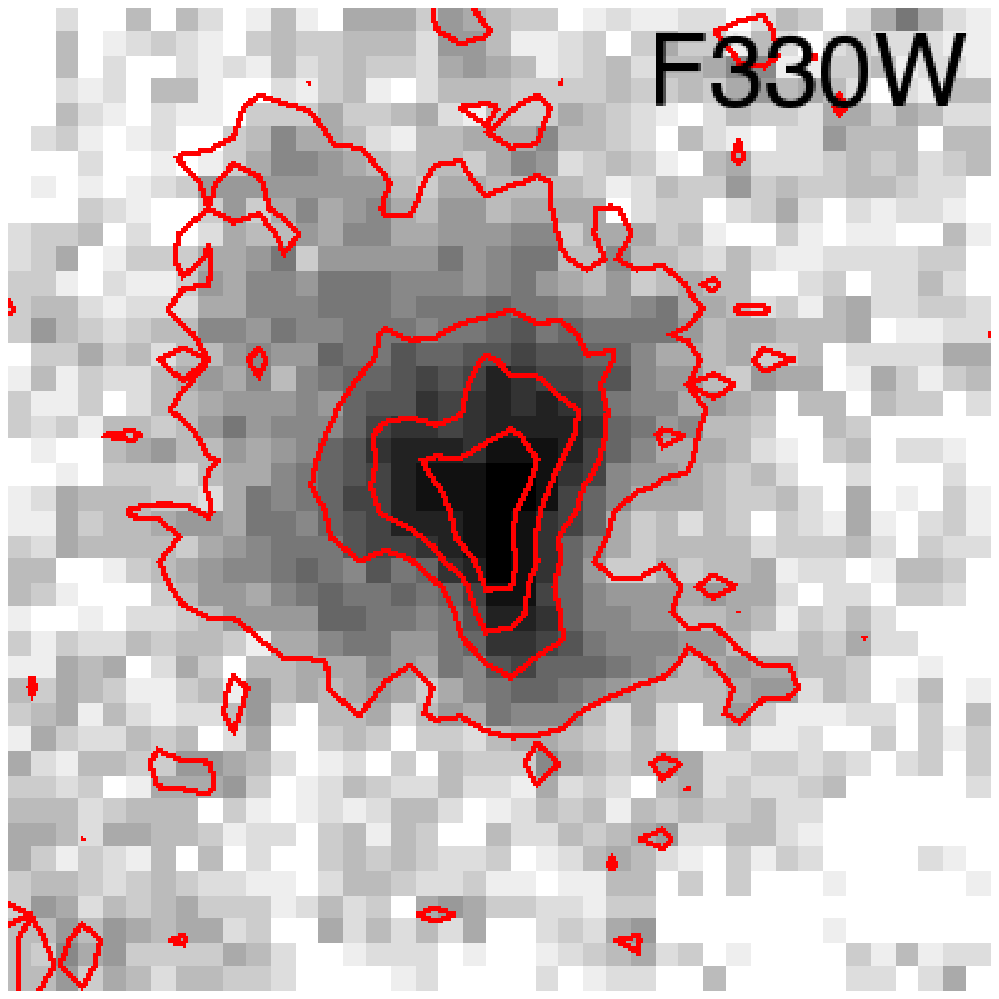}\includegraphics[width=0.24\textwidth]{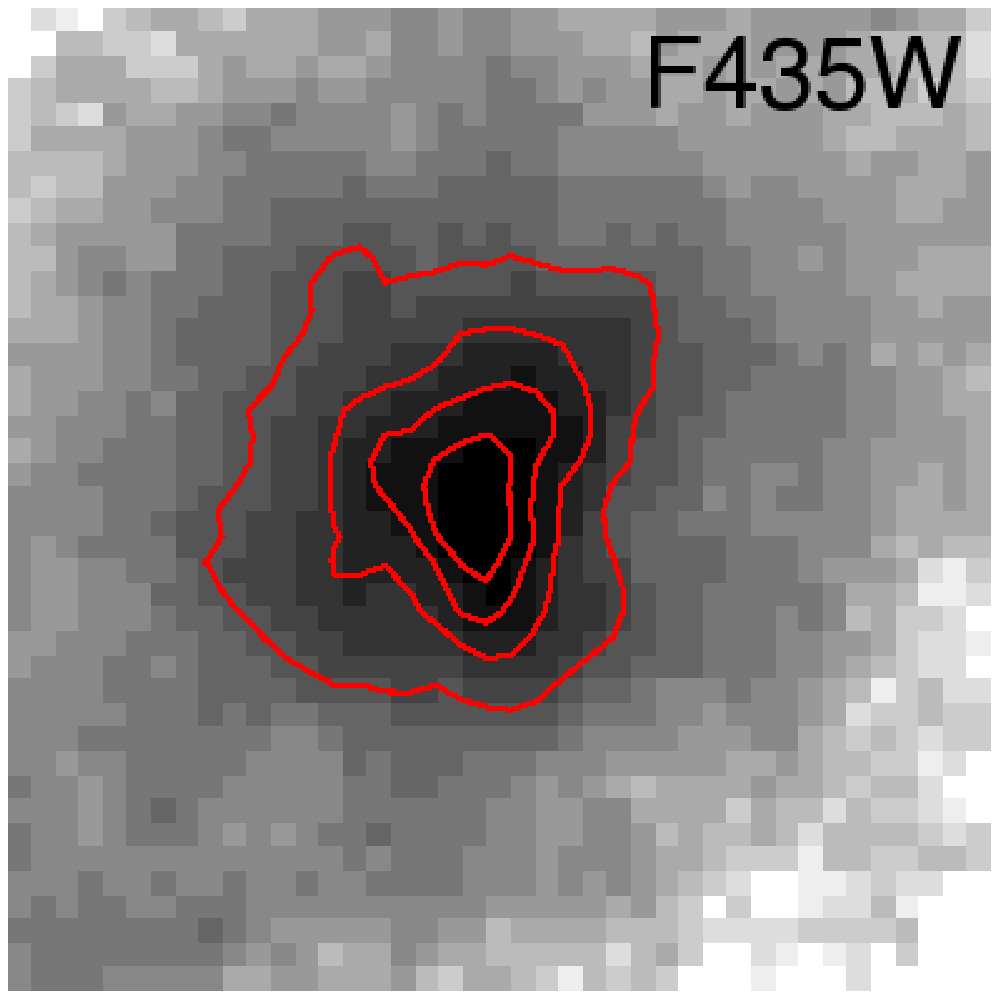}
\includegraphics[width=0.24\textwidth]{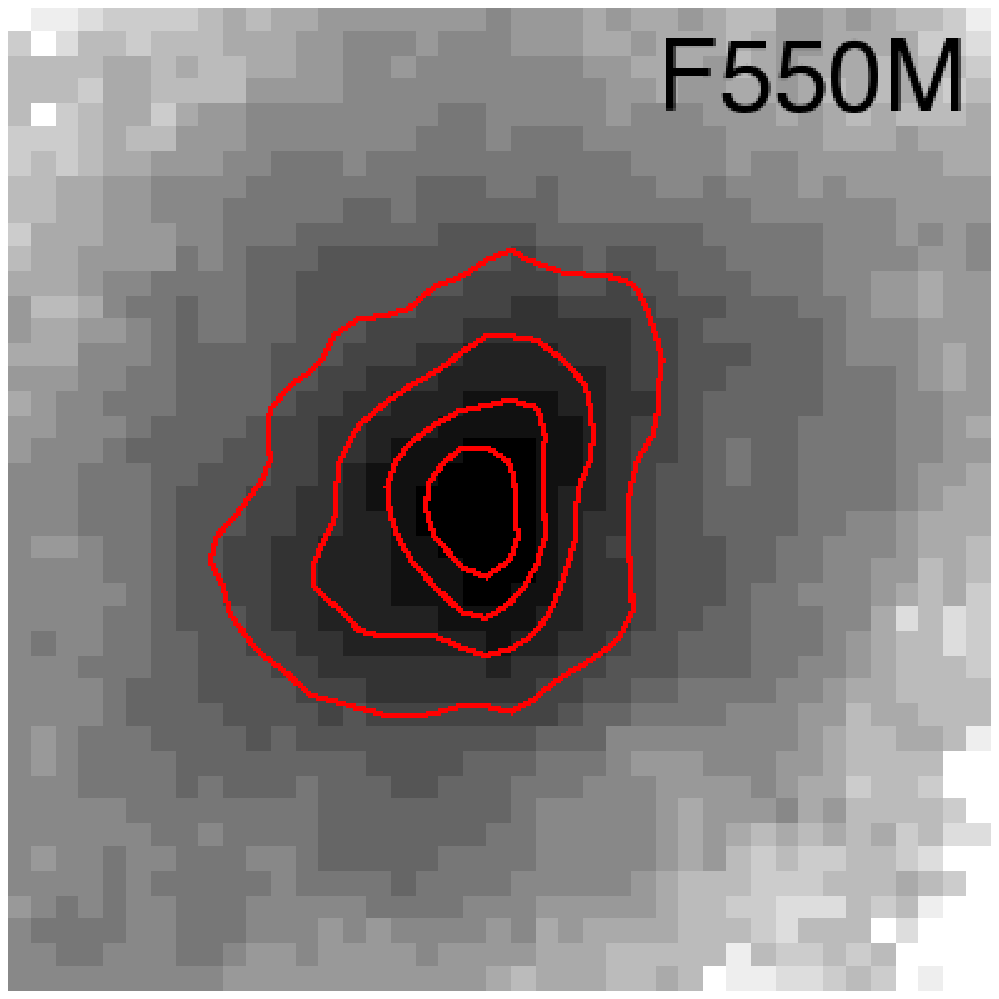}\includegraphics[width=0.24\textwidth]{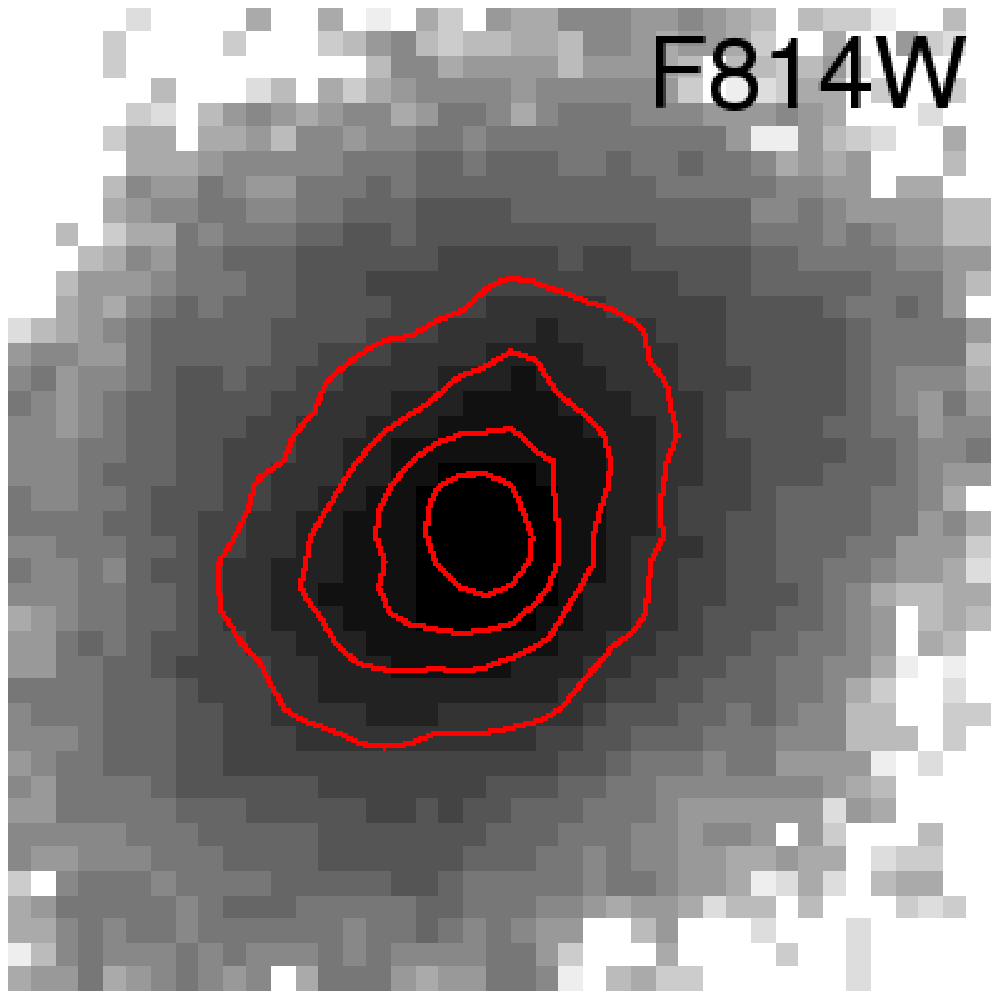}
\caption{
ACS/HRC $UBVI$ imaging of \#91 (M82-H). The top-left, top-right, bottom-left and bottom-right panels represent the F330W, F435W, F550M and F814W band images respectively, and the orientation follows that of the finding chart (Fig.~\ref{plate}), with East to the left and North to the top. The size of each image is 1$^{\prime \prime} \times$1$^{\prime \prime}$, which roughly corresponds to 18 parsecs at the distance of M82. The images are presented in a logarithmic intensity scale, while the contours show the linear scale flux levels at 20, 40, 60 and 80\% maximum intensity. Note how the $U$- and $B$-band images (F330W and F435W) show a highly asymmetric light profile, while in the $V$- and especially the $I$-band (F550M and F814W) the cluster light appears to be quite regular. Given that the extinction will have a greater effect on shorter wavelength bands, we can infer that the flux in the $U$ band is underestimated, hence causing the shift in colour-space that we observe in Fig.~\ref{fig:colplot}.
}
\label{h-plots}
\end{center}
\end{figure}

In \S~\ref{sec:ages_phot} we briefly discussed the fact that the age of cluster \#91 (M82-H) cannot be determined photometrically, as it lies beyond the area in colour-space covered by the SSP model evolutionary track, once extinction has been accounted for. This appears to be caused by an underestimation of its $U$-band flux, so it also affects the analytical photometric method applied (3DEF).

In support of this hypothesis, Fig.~\ref{h-plots} shows ACS-HRC imaging of the cluster in the $U$, $B$, $V$ and $I$ bands, obtained from the HST archive (program \#10609, PI Vacca), from which the cluster profile appears to be highly irregular in the $U$-band, while becoming smoother and more regular in the redder wavelength bands. This is the result of a complicated differential extinction pattern. Our measurements indicate that parts of the cluster are entirely obscured in the $U$ and $B$ bands, which can account for this shift in colour-space as it would spuriously raise the $U-B$ value.

\begin{figure}
\begin{center}
\includegraphics[width=0.48\textwidth]{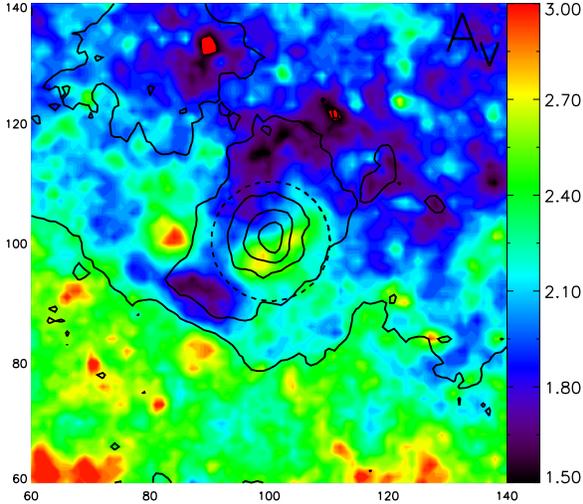}
\caption{Extinction map for cluster \#91 (M82-H). The orientation follows that of Figs.~\ref{plate} and \ref{h-plots} and the axes are given in arcseconds, with the x- and y-axis zeropoints placed roughly at the centre of the cluster. The solid lines show isophotes as calculated from the F435W flux and the dashed circle represents the aperture used in measuring the cluster magnitude. From this we deduce that the cluster is obscured by a dust lane that runs through the western edge of the cluster. This extinction seems to be disproportionate between different bands, hence the shift in colour-space we see in Fig \ref{fig:colplot}.}
\label{h-ext}
\end{center}
\end{figure}

In order to verify this inference, we map out the extinction on and around the cluster pixel-by-pixel in Fig.~\ref{h-ext}; the colour map represents the different levels of reddening, as determined by multi-band photometry and the solid lines represent observed isophotes in the F435W band. The technique we use is explained in Bastian et al.~(2007). In brief, we construct a colour map by measuring the flux of each pixel in the $B$, $V$ \& $I$ bands and therefore measure its reddening, while assuming the spectroscopically derived age. In principle, as we assume the cluster to be a SSP, we expect each pixel to have the same colour, therefore it is only the variation of extinction across the face of a cluster that causes the colour to change. The extinction map verifies the complexity of the reddening in the immediate vicinity of M82-H, showing a dust lane running through the eastern edge of the cluster. The location and physical characteristics of this extinction patch are consistent with the observed shape of the cluster in the $U$-band. For cluster \#91 we therefore use the age as derived by spectroscopy as the variable extinction should not change the spectroscopic features of the cluster.

We also find tentative evidence of a similar differential extinction pattern obscuring cluster \#108, however the S/N of this cluster is not high enough to quantitatively assess its effects. We are able to place the extinction range between 1.5 and 2.0 mags in $A_V$ in a manner that does not appear to be related with the underlying cluster structure. In addition, this cluster is not detected in the $U$-band, and this is the reason for its absence from the colour-colour plot (Fig~\ref{fig:colplot}).

\section{Discussion}\label{sec:disc}
In this paper we have presented an analysis of seven YMCs in region B of M82. Our results, along with age-dating of a further 29 clusters presented in \citet{phot} suggest that these clusters are consistent with being part of a young cluster population that started forming $\lesssim\,$220\,Myr ago. We find extinctions in the region to vary between 1.1--2.5\,mag and we also find differential extinction across the face of cluster \#91 and possibly \#108. These results largely contradict studies presented in the past and we discuss them in more detail in the rest of this section.

\begin{table}
\caption{Summary of extinction and age measurements and comparison to values published in \mbox{\citet[][denoted as G2003]{RdG03-1}}.}
\begin{center}
\begin{tabular}{lcccccc}
\tableline
\tableline
 & $A_V$ & $A_V^{\phantom{_V}{\rm G2003}}$ & $\tau_{\rm phot}$ & $\tau_{\rm spec}$ & $\tau_{\rm adopted}$ & $\tau_{\rm G2003}$\\
\# & \tmult{(magnitudes)} & \multicolumn{4}{c}{(Myr)}\\
\tableline
91	& 2.4		& 0.00	& \ldots			& 180	& 180			& 5800\\
97	& 1.1 	& 0.06	& 180			& 190	& 190 			& \phantom{1}580\\
103	& 1.9 	& 0.12	& \phantom{1}90	& 100	& 100			& 1000\\
108	& 2.5 	& 0.50\tablenotemark{a} & \ldots & 140	& 140			& 3200\tablenotemark{a}\\
113	& 1.9 	& 1.60	& 170			& 100	& 100			& \phantom{1-}3.8\\
126	& 1.5 	& 0.00	& 260			& 200	& 200			& 1400\\
131~~ & 1.2 	& 0.12	& 170	& \phantom{1}80	& \phantom{1}80	& \phantom{1}720\\
\tableline
\end{tabular}
\end{center}
\tablenotetext{a}{values taken from \citet{RdG01}, no data available in \citet{RdG03-1}}
\label{tab-summ}
\end{table}

\subsection{Extinction in region~B}
\label{sec:res-av}
M82 is a gas and dust rich galaxy, which is seen nearly edge on, causing its stellar population to be largely hidden in the optical. Using optical/near-UV photometry \citep{phot} and spectroscopy, we find that the $A_V$ extinction for clusters in our sample shows a wide range, with evidence of other clusters being more highly reddened. For example, cluster \#108 has the highest extinction of 2.5 mag and is not visible in the $U$-band. This variable extinction suggests that region B presents a line of sight into the galaxy of relatively low extinction, which therefore allows us to probe depths that are not visible on the western side of the galaxy, similar to a ``Baade's window'' in the Milky Way. \citet{smith01} find a similar patch of high, variable extinction in the vicinity of cluster M82-F, measured by \citet{bastian07} to have $A_V$ values between 2--4~mag, with much higher extinction values in the surrounding region.

The study by de Grijs et al. (2003a) concluded that the clusters in that sample were observable because they suffered from very little extinction ($A_V\lesssim1.2\,$mag). They suggest that all the clusters are situated near the surface of region~B with respect to our line of sight, implying that the region must have an extremely steep gas/dust density gradient. In contrast, we find a larger range in extinctions and a variable overall extinction along this line of sight and hence on average lower compared to other regions of the galaxy disk.

The selection effect  created by extended areas of variable extinction reaching low values may explain why region B was thought to occupy a special place within M82, as it allows for the observation of a far  larger number of clusters in visible light bands. A high number of clusters could be interpreted as the result of a higher cluster formation rate in the region. The interpretation offered in this paper is that these `holes' in the dust distribution of the galaxy are in fact the source of the discrepancy in appearance between its western and eastern sides. This claim is supported by NICMOS near IR observations of the galaxy \citep{alonso01,alonso03}, where the two sides seem to have comparable luminosity and to host a similar number of clusters, indicating a shared cluster formation history.

\subsection{Cluster ages}
\label{res-age}
We find that the spectroscopic ages of the seven clusters we have analysed in region B are between 80 and 200\,Myr. Comparison of the photometric and spectroscopic ages for the five clusters in common show that overall, the preferred spectroscopic ages are towards the younger part of the allowed photometric age range.  A good example of this is cluster \#126 which has a photometric age range of 200--780\,Myr (Fig. 3, Table 2) but the spectroscopy results clearly narrow this down to 160--280\,Myr, and the spectrum shows that this cluster cannot be as old as $800\,$Myr.

We now compare our ages with the ages of de Grijs et al. (2003a), as given in Table 4. It is clear that, with the exception of \#113, the de Grijs et al. ages are considerably older. In the larger photometric sample of 35 clusters presented by \citet{phot}, the ages range from 8--310~Myr\footnote{This applies to 33 of the 35 clusters in this sample. The remaining clusters \#16 and 35 appear to have ages of 2.3 and 1.3\,Gyr respectively, with minimum acceptable values of 130 and 107\,Myr. These two values are poorly constrained and therefore not representative of the overall age distribution.} and the peak is at 150 Myr. Overall, we conclude that the region B cluster sample is much younger than presented in \citet{RdG01, RdG03-1}.

The ages that we find for the clusters agree well with the time-scale of the last encounter 
between M82 and M81, based on a value of 220\,Myr (Yun 1999). de Grijs et al. (2001) also suggest that the age of the region B cluster population is consistent with this encounter but they use an earlier age estimate by \citet{brouillet91} of 510\,Myr.

\subsection{Radial Velocities}\label{sec:res-rvs} 
It has been suggested by \citet{RdG03-1} that region~B is a gravitationally bound structure. In order to explain how such a large structure could form and survive despite the effects of differential rotation, they suggested that the presence of the stellar bar \citep{wills00} may have caused the formation of a circumnuclear ring, of which region B is part. It is well known that a ring or torus does indeed exist  around the M82 bar, but we note that if the torus were to include region B, it would have to have a width of $\sim$500\,pc, in clear disagreement with the observations \citep[$r=$86\,pc from the centre according to][Wei\ss\ et al. 2001 find a radius for the molecular torus of $r=\,$65\,pc]{achtermann95}.

\begin{figure}
\begin{center}
\includegraphics[width=0.49\textwidth]{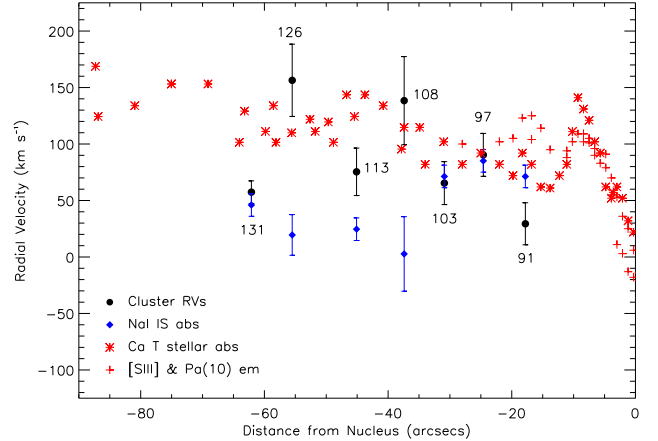}
\caption{Major-axis position-velocity diagram for the eastern half of the disk. The cluster measurements are labelled with their respective IDs. The Ca\,T stellar absorption and [S\three] and Pa(10) emission measurements are taken from \citet{mckeith93}.}
\label{rot-curve}
\end{center}
\end{figure}

In order to explore the dynamical state of the region~B cluster sample, we plot their derived radial velocities and the corresponding Na\one\ interstellar absorption line measurements in Fig.~\ref{rot-curve}. We have also included near infrared Ca\two\,$\lambda$8498,8542,8662 (Ca~T) stellar absorption and [S\three]$\lambda$6531,9069 and Pa(10) nebular emission line measurements from \citet{mckeith93}.

Clusters \#97 and 103 exhibit stellar and gas kinematics that are consistent with the galaxy rotation curve traced by the Ca~T velocities, and the similarity between the cluster and Na\one\ measurements shows that the interstellar gas is located near the clusters. The radial velocities of clusters \#108 and 126 are consistent with the Ca~T absorption line measurements and show that they are located within the main stellar disk of M82. However, the corresponding Na\one\ velocities are near to the systemic value, indicating that a large column of cool IS gas is located in front of these clusters, on the outskirts of the disk. The radial velocities of clusters \#91, 113 and 131 are consistent with them being located on the far- or near-side of the disk. The fact that the Na\one\ IS lines are close to systemic argues, however, that these three clusters must be on the near-side of the disk otherwise we would expect the Na\one\ velocity to be offset because of the greater path length.

We thus find no indication that our sample of clusters has unusual kinematics; instead the large range in radial velocities is consistent with them being located at different depths within the M82 disk. This also explains the large variation in extinction that we measure.

Our findings allow us to test the hypothesis that region B is bound. In order for this to be true, the clusters should trace equal angles in equal amounts of time as they orbit about the galaxy centre. The galactocentric distances of these seven clusters places them on the gently rising part of the rotation curve. Assuming a constant rotational velocity $v_\mathrm{rot}=$120~\kms\ (Fig.~\ref{rot-curve}), this corresponds to an orbital period of 18 and 54\,Myr for the inner and outer boundaries of the region (assumed to be marked by clusters \#91 and 131 respectively that are situated 700\,pc apart). This would imply that clusters in the outermost regions should have a radial velocity, $v_\mathrm{rad}$, three times that of the ones in the innermost regions. This is clearly not the case; we find no correlation between the cluster radial velocities and their spatial positions as they seem to scatter about the 120\kms\ mark, and conclude that the clusters are on regular orbits with respect to the stellar component of the disk.

Finally, we remark that through a number of arguments, \citet{RdG03-1} found that the cluster population of region B has a log-normal mass function. This supported their hypothesis that region B is bound, since a normal power law function would imply unphysically high initial densities.

\citet{phot} suggest that the turn-over in the cluster luminosity and mass functions is not real but is caused by the neglect of the effect of extended sources on the detection limit. This, in addition to the younger ages that we find, suggests that the disruption timescale derived in \citet{RdG03-1} and used in \citet{RdG05} is probably too short. By not imposing a short disruption timescale to such a young cluster population, the idea of an initial power-law distribution is consistent with the data.

\section{Conclusions}
We have presented Gemini GMOS spectroscopy and ACS \emph{UBVI} photometry for seven clusters in region~B of the starbust galaxy M82. Our aims were to obtain accurate ages, extinctions and radial velocities for the clusters.

We have used both photometric and spectroscopic methods to determine cluster ages. Our main age determination method is fitting the Balmer spectral lines with all available models and finding the best fit (i.e. the lowest $\chi^2$ on the overall line profile fit), in order to eliminate the known Balmer line strength degeneracy. We find our method to agree with photometric ages derived using the 3DEF method (based on $UBVI$ colours). This demonstrates that the photometrically-derived ages are accurate but they are not as precise as the ones obtained from spectroscopy and confirms that the only truly independently reliable method for age-dating clusters in extragalactic environments is spectroscopy. The inclusion of $U$-band photometry may enhance the accuracy of photometric measurements, but unless they can be cross-checked by spectroscopic means, the potential age degeneracy may not be broken. 

We find cluster ages in the range of 80 to 200\,Myr, a distribution that is consistent with the timescale for the last encounter between M82 and M81 as proposed by \citet{yun99}, namely 220\,Myr. These findings in combination with the larger photometric sample of \citet{phot}, disagree with the `fossil starburst' scenario proposed by \citet{RdG01}, where region B in M82 was identified with the remnant of an off-centre starburst that commenced about 600\,Myr ago, a time scale based on their photometric determinations of star cluster ages. We suggest that the increased cluster formation rate in region~B is representative of the era of increased star formation across the galaxy disk triggered by the last encounter with M81.

The extinction along our line of sight in this extended region appears to vary greatly, between 1.1 and 2.5 magnitudes. As M82 is seen virtually edge-on, we find that \mbox{M82-B} presents a view into various depths of the body of the galaxy through an arrangement of `windows' in the dust distribution \citep[as hypothesised for cluster M82-F by][]{smith01}. We also find differential extinction across the face of cluster \#91 (and also possibly \#108), which sets a serious obstacle in the photometric determination of extinction and age in these cases. In fact, the effect of dust in this environment is in some cases so grave that photometry cannot be used at all as an age/extinction diagnostic.

We have also used the available spectroscopy to derive cluster kinematics. This allows us to reinforce our extinction-based findings and show that the seven clusters reside at different depths within the disk of M82. The large scatter of cluster velocities about the gently rising component of the rotation curve indicates that the clusters do not move in a co-ordinated fashion and that region B cannot be bound.

Previous studies of clusters F and its neighbour L on the western side of the disk show that they both have an age of $60\pm20$ Myr \citep{GS99,bastian07}. This age fits in well with the region B cluster age distribution presented in this paper and \citet{phot}, and supports our suggestion that the increase in the cluster formation rate was not local to region B but part of a galaxy-wide burst.
In addition, the very high masses of clusters F \citep[$\sim10^6$\,\Msun, ][]{smith01,bastian07} and L \citep[$4\times10^6$~\Msun, ][]{mccrady07} imply that many lower mass clusters were also likely to have formed with F and L.

In summary, we find region~B to be optically bright owing to the presence of low internal extinction patches, thus offering a deep view into the M82 disk at radii between $\sim$ 400--1200\,pc. The range of cluster ages and other properties in this region then are typical of the evolution of the main body of M82 and reflect  the large increase in star formation that occurred about 220\,Myr ago when M82 last passed close to M81. 
M82 region B stands out because it is representative of the mid-disk zone, and is relatively clear of dust, rather than being a special substructure.  This model also receives support from the larger sample of photometric ages in Smith et al. (2007), and will be further discussed in the Konstantopoulos et al. (in preparation) study of another three dozen M82 star clusters distributed across M82.
\\
\acknowledgments
We thank the staff at Gemini North for obtaining the spectroscopic observations on which this paper is based.
We also thank the Hubble Heritage team at the Space Telescope Science Institute for the mosaic image of M82.
JSG appreciates support for studies of starburst galaxies provided by the University of Wisconsin-Madison Graduate School.
ISK would like to acknowledge the support of the Perren Fund, provided by the Astrophysics Group at University College London.
Support for program \#10853 was provided by NASA through a grant from the Space Telescope Science Institute, which is operated by the Association of Universities for Research in Astronomy, Inc., under NASA contract NAS 5-26555.




\begin{thebibliography}{99}

\bibitem[Achtermann \& Lacy (1995)]{achtermann95} Achtermann, J.~M. \& Lacy, J.~H., 1995, \apj, 439, 163

\bibitem[Alonso-Herrero et al.(2001)]{alonso01} Alonso-Herrero, A., Rieke, M.~J., Rieke, G.~H., \& Kelly, D.~M.\ 2001, \apss, 276, 1109 

\bibitem[Alonso-Herrero et al.(2003)]{alonso03} Alonso-Herrero, A., Rieke, G.~H., Rieke, M.~J., \& Kelly, D.~M.\ 2003, \aj, 125, 1210 

\bibitem[Anders \& Fritze-v. Alvensleben (2003)]{anders03} Anders, P., \& Fritze-v. Alvensleben., U. 2003, A\&A, 401, 1063

\bibitem[Anders et al.(2004)]{anders04} Anders, P., Bissantz, N., Fritze-v.~Alvensleben, U., \& de Grijs, R.\ 2004, \mnras, 347, 196 

\bibitem[Bastian et al.(2005)]{bastian05} Bastian, N., Gieles, M., Lamers, H.~J.~G.~L.~M., Scheepmaker, R.~A., \& de Grijs, R.\ 2005, A\&A, 431, 905 

\bibitem[Bastian et al.(2007)]{bastian07} Bastian, N., Konstantopoulos, I.~S., Smith, L.~J., Westmoquette, M.~S., Trancho, G., \& Gallagher, J.~S., III, 2007, \mnras, 379, 1333

\bibitem[Bik et al.(2003)]{bik03} Bik, A., Lamers, H.~J.~G.~L.~M., Bastian, N., Panagia, N., \& Romaniello, M.\ 2003, A\&A, 397, 473 

\bibitem[Brouillet et al.(1991)]{brouillet91} Brouillet, N., Baudry, A., Combes, F., Kaufman, M., \& Bash, F.\ 1991, \aap, 242, 35 

\bibitem[Bruzual \& Charlot(2003)]{bc03} Bruzual, G., \& Charlot, S.\ 2003, \mnras, 344, 1000 

\bibitem[Calzetti (1997)]{calzetti97} Calzetti, D.\ 1997, \aj, 113, 162
 
\bibitem[Cappellari \& Emsellem(2004)]{ppxf} Cappellari M., \& Emsellem E., 2004, PASP, 116, 138.

\bibitem[de Grijs et al.(2001)]{RdG01} de Grijs, R., O'Connell, R.~W., \& Gallagher, J.~S., III, 2001, \aj, 121, 768

\bibitem[de Grijs et al.(2003a)]{RdG03-1} de Grijs, R., Bastian, N., \& Lamers, H.~J.~G.~L.~M.\ 2003a, \mnras, 340, 197

\bibitem[de Grijs et al.(2003b)]{RdG03-2} de Grijs, R., Bastian, N., \& Lamers, H.~J.~G.~L.~M.\ 2003b, \apj, 583, L17

\bibitem[de Grijs, Parmentier, \&~Lamers(2005)]{RdG05} de Grijs, R., Parmentier, G., \& Lamers, H.~J.~G.~L.~M.\ 2005, \mnras, 364, 1054

\bibitem[Fall \& Zhang (2001)]{fall01} Fall, S.~M., \& Zhang, Q.\ 2001, \apj, 561, 751 

\bibitem[Filippenko(1982)]{filippenko} Filippenko, A.~V.\ 1982, \pasp, 94, 715 

\bibitem[Gallagher \& Smith(1999)]{GS99} Gallagher, J.~S., \& Smith, L.~J.\ 1999, \mnras, 304, 540 

\bibitem[Gieles et al.(2006a)]{gieles06a} Gieles, M., Larsen, S.~S., Bastian, N., \& Stein, I.~T.\ 2006a, \aap, 450, 129 

\bibitem[Gieles et al.(2006b)]{gieles06b} Gieles, M., Portegies Zwart, S.~F., Baumgardt, H., Athanassoula, E., Lamers, H.~J.~G.~L.~M., Sipior, M., \& Leenaarts, J.\ 2006b, \mnras, 371, 793 

\bibitem[Gonz\'alez Delgado et al.(2005)]{gd05} Gonz\'alez Delgado, R.~M.~G., Cervi{\~n}o, M., Martins, L.~P., Leitherer, C., \& Hauschildt, P.~H.\ 2005, \mnras, 357, 945 

\bibitem[Goudfrooij et al.(2004)]{goud1316} Goudfrooij, P., Gilmore, D., Whitmore, B.~C., \& Schweizer, F.\ 2004, \apjl, 613, L121 

\bibitem[Goudfrooij et al.(2007)]{goud3610} Goudfrooij, P., Schweizer, F., Gilmore, D., \& Whitmore, B.~C.\ 2007, \aj, 133, 2737 

\bibitem[Larsen(2004)]{larsen04} Larsen, S.~S.\ 2004, \aap, 416, 537 

\bibitem[Marcum \& O'Connell (1996)]{marcum96} Marcum, P.~M., \& O'Connell, R.~W.\ 1996, From Stars to Galaxies: the Impact of Stellar Physics on Galaxy Evolution, 98, 419 

\bibitem[McCrady \& Graham(2007)]{mccrady07} McCrady, N., \& Graham, J.~R.\ 2007, ApJ, 663, 844

\bibitem[McKeith et al.(1993)]{mckeith93} McKeith, C.~D., Castles, J., Greve, A.,~\& Downes, D.~1993, A\&A, 272, 98

\bibitem[McLeod et al.(1993)]{mcleod93} McLeod, K.~K., Rieke, G.~H., Rieke, M.J., ~\& Kelly, D.~M.~1993, ApJ, 412, 111

\bibitem[Mutchler et al.(2007)]{mosaic} Mutchler et al., 2007, PASP, 119, 1

\bibitem[O'Connell \& Mangano (1978)]{om78} O'Connell R.~W., \& Mangano J.~J., 1978, \apj, 221, 62

\bibitem[Savage \& Mathis (1979)]{savage79} Savage, B.~D. \& Mathis, J.D. 1979, ARAA, 17, 73

\bibitem[Sirianni et al.(2005)]{sirianni05}  Sirianni, M., Jee, M.~J., Ben{\'{\i}}tez, N., et al.~2005, PASP, 117, 1049

\bibitem[Smith \& Gallagher (2001)]{smith01} Smith, L.~J., \& Gallagher, J.~S.\ 2001, \mnras, 326, 1027 

\bibitem[Smith et al.(2006)]{stis} Smith, L.J., Westmoquette, M.S., Gallagher, J.S., III, O'Connell, R.W., Rosario, D.J., \& de Grijs, R. 2006, \mnras, 370, 513

\bibitem[Smith et al.(2007)]{phot} Smith, L.J., Bastian, N., Konstantopoulos, I. S., Gallagher, J.S., Gieles, M., de Grijs, R., Larsen, S.S., O'Connell, R.W., \& Westmoquette, M.S., 2007, ApJL, 667, L145

\bibitem[Trancho et al.(2007a)]{gelys07a} Trancho, G., Bastian, N., Schweizer, F., \& Miller, B.~W.\ 2007, \apj, 658, 993

\bibitem[Trancho et al.(2007b)]{gelys07b} Trancho, G., Bastian, N., Miller, B.~W., \& Schweizer, F.\ 2007, ApJ, 664, 284

\bibitem[Vesperini \& Zepf(2003)]{vesperini03} Vesperini, E., \& Zepf, S.~E.\ 2003, \apjl, 587, L97 

\bibitem[Walborn(1972)]{walborn72} Walborn, N.~R.\ 1972, \aj, 77, 312 

\bibitem[Walborn \& Fitzpatrick(1990)]{walborn90} Walborn, N.~R., \& Fitzpatrick, E.~L.\ 1990, \pasp, 102, 379 

\bibitem[Wills et al.(2000)]{wills00} Wills, K.~A., Das, M., Pedlar, A., Muxlow, T.~W.~B., Robinson, T.~G., 2000, MNRAS, 316, 33 

\bibitem[Yun (1999)]{yun99} Yun, M.S., 1999, in: IAU Symp. 186, Galaxy Interactions at Low and High Redshift, ed. J.~E. Barnes \& D.~B. Sanders (Dordrecht: Kluwer), 81

\end{thebibliography}
\end{document}